\documentclass[aps,twocolumn,superscriptaddress,a4paper]{revtex4}
\usepackage{amsfonts,amssymb,amsmath}
\usepackage[utf8]{inputenc}

\usepackage[pdftex,breaklinks=true,colorlinks=true]{hyperref}
\usepackage[pdftex]{graphicx}
\usepackage{leftidx}

\DeclareMathOperator{\Tr}{Tr}

\newcommand{\ra}{\rangle}

\newcommand{\ket}[1]{\left|#1\right>}

\begin{document}

\title{Finite-size scaling of the Shannon-R\'enyi entropy in two-dimensional systems with spontaneously broken continuous symmetry}

\author{Gr\'{e}goire Misguich}
\affiliation{Institut de Physique Théorique, Université Paris Saclay, CEA, CNRS, F-91191 Gif-sur-Yvette, France}
\author{Vincent Pasquier}
\affiliation{Institut de Physique Théorique, Université Paris Saclay, CEA, CNRS, F-91191 Gif-sur-Yvette, France}
\author{Masaki Oshikawa}
\affiliation{Institute for Solid State Physics, University of Tokyo, Kashiwa 277-8581, Japan}

\begin{abstract}
We study the scaling of the (basis dependent) Shannon entropy for
two-dimensional quantum antiferromagnets with N\'eel long-range
order. We use a massless free-field description of the gapless spin wave
modes and phase space arguments to treat the fact that the finite-size
ground state is rotationally symmetric, while there are degenerate
physical ground states which break the symmetry.  Our results show that
the Shannon entropy (and its Rényi generalizations) possesses some
universal logarithmic term proportional to the number $N_\text{NG}$ of
Nambu-Goldstone modes. In the case of a torus, we show that $S_{n>1} \simeq
{\rm const.} N+ \frac{N_\text{NG}}{4}\frac{n}{n-1} \ln{N}$
and 
$S_1 \simeq 
{\rm const.} N - \frac{N_\text{NG}}{4} \ln{N}$,
where $N$
is the total number of sites and $n$ the Rényi index. The result
for $n>1$ is 
in reasonable agreement with
the quantum Monte Carlo results of Luitz {\it et al.}
[Phys. Rev. Lett. 112, 057203 (2014)], and qualitatively similar to
those obtained previously for the {\it entanglement} entropy.  The
Shannon entropy of a line subsystem (embedded in the two-dimensional
system) is also considered.  Finally, we present some density-matrix
renormalization group (DMRG) calculations for a spin$\frac{1}{2}$ XY
model on the square lattice in a cylinder geometry.  These numerical
data confirm our findings for logarithmic terms in the $n=\infty$ Rényi
entropy (also called $-\ln{p_{\rm max}}$).  They also reveal some universal
dependence on the cylinder aspect ratio, in good agreement with the fact
that, in that case, $p_{\rm max}$ is related to a non-compact free-boson
partition function in dimension 1+1.
\end{abstract}

\maketitle

\section{Introduction}

Recently, there is a growing interest in utilizing
information-theoretic quantities to characterize phases,
to go beyond the traditional characterizations based on
order parameters and correlation functions.
The most popular among them is entanglement entropy.
In fact, many low-energy and long-distance properties of quantum
many-body systems can be extracted from the scaling of the entanglement
entropy of some large subsystem. Two prototypical examples are critical
spin chains, where the central charge can be read off from the scaling
of the entanglement of a
segment~\cite{Holzhey_geometric_1994,vidal_entanglement_2003,calabrese_entanglement_2004},
and that of gapped topologically ordered states in dimension two, which have some
universal subleading contributions related to the the nature of the
fractionalized excitations (quantum dimensions) of the
phase~\cite{kitaev_topological_2006,levin_detecting_2006}.

It was also realized that a somewhat simpler entropy, the (basis-dependent) Shannon entropy, share some similar properties (see \cite{luitz_participation_2014} for a review).
It is defined as follows.
When expanded in some discrete basis $\left\{\left|i\right>\right\}$,
a quantum state $\left|\psi\right>$ defines a set 
of probabilities
\begin{equation}
 p_i= \left| \left<\psi | i \right> \right|^2
 \label{eq:pi}
\end{equation}
that can, in turn, be used to define a Shannon entropy:
\begin{equation}
 S_1 = - \sum_i p_i \ln{p_i}.
\end{equation}
In the following it will also be useful to consider a generalization of this entropy, the Shannon-Rényi entropy (SRE):
\begin{equation}
S_n =\frac{1}{1-n}\ln{\left(\sum_i p_i^n\right)},
 \label{eq:SREdef}
\end{equation}
with Eq.~\ref{eq:pi}.
For one-particle problems  described by a wave function $\psi(r)$ in real space, the entropies are simply related by $S_n = \frac{1}{1-n} \ln{P_n}$ to the
so-called inverse participation ratios~: $P_n=\int d^D \mathbf{r} \left|\psi(r)\right|^{2n}$. 
The latter
measure how spatially localized is the particle. In the presence of disorder it can be used to detect  Anderson metal-insulator
transitions~\cite{evers_anderson_2008}, as follows:
in the delocalized phase one has $P_n\simeq L^{-D (n-1)}$, where $D$ is the spatial dimension, and
$L$ the linear system size (thus $L^D$ is the Hilbert space dimension).
In contrast, one has instead $P_n\simeq L^0$ in the localized phase. 
At the transition point it scales like $P_n\simeq L^{-\alpha_n (n-1)}$
where $\alpha_n$ defines a continuous family of critical exponents
(multifractality).
In the notation of Eq.~\ref{eq:SREdef} it means
$S_n\simeq \alpha_n \ln{L}$.

The situation is quite different for many-body systems, which
have exponentially many ($\propto \exp{\left[ \mbox{const.} L^D \right]}$)
basis states.
A generic many-body wave function has some nonzero weights
on a finite fraction of these basis states.
As a consequence, the leading behavior of the SRE
(Eq.~\ref{eq:SREdef}) is generically
a {\em volume law}, which means $S_n\simeq \alpha_n L^D$.
This may also be interpreted as ``multifractality''~\cite{atas_multifractality_2012} with interesting features at
phase transitions~\cite{monthus_pure_2015}.
However, it should be noted that, contrary to that in one-body wave
functions, the ``multifractality'' is generic for wave functions
living in a many-body Hilbert space, even for featureless product
states.  Consider for instance $N$ independent spin-$\frac{1}{2}$ in the
same state $\cos(\theta)|\uparrow\ra+\sin(\theta)|\downarrow\ra$.  The
SRE of that tensor product state is
$S_n=\frac{N}{1-n}\ln\left(\cos^{2n}{\theta} +\sin^{2n}{\theta} \right)$,
which is a nonlinear function of $n$.
Here the number of spins, $N$, corresponds to the volume $L^D$ of
the system.
In fact, the coefficient of the leading volume-law term in
the SRE is generally non-universal and depends on microscopic details,
as it is evident in the above simple example.
Thus it is not an interesting quantity from the viewpoint of
elucidation of universal behavior in a quantum phase.
Nevertheless, subleading terms can contain universal information
which are determined by the long-distance properties of the system.
This has been studied in quantum spin chains in
particular, and these corrections are typically $\mathcal{O}(1)$ for
periodic
chains~\cite{stephan_shannon_2009,stephan_renyi_2010,kumano_renyified_2015},
and $\mathcal{O}(\ln L)$ for open chains
\cite{zaletel_logarithmic_2011,stephan_phase_2011}.

In this paper we are interested in two-dimensional (2D) quantum
antiferromagnets, where the spin rotation symmetry -- U(1) or SU(2)
-- is spontaneously broken at zero temperature in the thermodynamic
limit.  In such systems with magnetic long-range order and gapless
Nambu-Goldstone modes, it was observed, using (modified) spin-wave
calculations~\cite{song_entanglement_2011} and quantum Monte Carlo (QMC)
on Heisenberg models~\cite{kallin_anomalies_2011} that the {\em
entanglement} entropy possesses some additive $\ln L$ corrections to
the boundary law. Soon after, these results were explained by some
analytical calculations (quantum rotor model and a nonlinear sigma
model)~\cite{metlitski_entanglement_2011}, leading to the prediction
that, in two dimensions, the coefficient of $\ln L$ is $N_\text{NG}/2$,
where $N_\text{NG}$ is the number of Nambu-Goldstone modes.
We note that a related discussion was also made for Heisenberg
ferromagnets in Refs.~\cite{popkov_logarithmic_2005,popkov_entangling_2005, castro-alvaredo_entanglement_2012}.
The situation is however quite different from antiferromagnets of our interest,
which have linearly dispersing Nambu-Goldstone modes
and a unique ground state in a generic finite-size system.

Recently, the SRE of several 2D magnets were computed using QMC~\cite{luitz_universal_2014,luitz_shannon-renyi_2014,luitz_participation_2014}.
By simulating spin-$\frac{1}{2}$ XXZ and Heisenberg models, U(1) and SU(2) broken symmetries were investigated.
The SRE for the complete system (torus), as well as the entropy of a line subsystem were measured.
In these studies, the basis states $|i\rangle$ chosen to define the probabilities $p_i$ (and  the SRE)
are eigenstates of the local magnetization $S^x(r)$. This basis choice  requires to select 
a particular spin direction in the $xy$ (easy) plane of the system, and such a choice therefore explicitly breaks the spin rotation
symmetry about the $z$ axis. In the present work we will also focus on such a situation, where the quantization axis
used to define the local basis is not invariant under the rotation symmetry of the Hamiltonian, but corresponds to a possible ordering direction for the order parameter ({\it i.e.} the sublattice magnetization).

In all the cases studied in Refs.~\cite{luitz_universal_2014,luitz_shannon-renyi_2014,luitz_participation_2014}, 
some additive $\ln L$ corrections were observed in the SRE in presence of magnetic long-range order.
Some of these results are summarized in Tab.~\ref{tab:toulouse}.
Motivated by these numerical results, 
 we study in this paper the SRE of these systems by using an effective relativistic (free boson) field theory
of the Nambu-Goldstone modes.
While we are primarily interested in the cases with
a spontaneously broken U(1) or SU(2) symmetry,
our analysis can be applied to the cases with
a more general spontaneously broken continuous symmetry.

It should be noted that, Nambu-Goldstone modes, which
accompany a spontaneous breaking of a continuous symmetry,
are classified into two categories: type-I and
type-II~\cite{Nielsen-Chada_NPB1976} or
type-A and type-B~\cite{Watanabe-Murayama_PRL2012}.
In this paper, we focus on the cases only with
the type-I (type-A with a linear dispersion) Nambu-Goldstone
modes, which can be described by the relativistic
free boson field theory.
In such cases, we can identify the number of the Nambu-Goldstone
modes $N_\text{NG}$ with the number of broken symmetry
generators.
We leave the analysis of the cases with type-II or type-B
modes to the future,
although some part of our discussion could be applied to these
cases as well.

We find a universal logarithmic term in the SRE
with respect to the system size, governed by
the number of modes $N_\text{NG}$.
Our theory is consistent with the numerical results
obtained by the Toulouse group, even though the quantitative agreement
is not perfect. 
We will also provide new numerical results for the SRE on
cylinders, to be compared with the theory.
We believe that our approach is on the right
track and could be extended for further quantitative improvements.

The paper is organized as follows.
In Sec.~\ref{sec:osc}, we analyze the contribution to the SRE of the 
fluctuations due to Nambu-Goldstone modes.
We first focus on the $n=\infty$ limit of the SRE,
$S_{\infty} \sim -\ln p_{\rm max}$, where $p_{\rm max}$ is
the largest among the probabilities of finding a particular basis
configuration upon the corresponding projective measurement
of the ground state.
As far as the universal terms are concerned, we show that this
problem is closely related to the determinant of the Laplacian in 2D
(Sec.~\ref{ssec:Laplacian}).
While we find a universal logarithmic term, its coefficient has
the opposite sign to that obtained with QMC.
The discrepancy is attributed to the degeneracy of the ground states
in the presence of spontaneous breaking of a continuous symmetry,
as discussed in Sec.~\ref{sec:TOS}.
In Sec.~\ref{sec:ndep}, we combine the results from earlier sections
to derive the universal logarithmic term in SRE for $n>1$ and $n=1$.
In Sec.~\ref{sec:line} we also discuss the logarithmic terms
in the SRE of a subsystem which has the geometry of a straight line embedded in a 2D system, for which Luitz {\it et al.}~\cite{luitz_shannon-renyi_2014,luitz_quantum_2016}
have some QMC data indicating clearly the presence of universal log terms.
Sec.~\ref{sec:dmrg} presents some 2D DMRG calculations
of the ground state of the spin-$\frac{1}{2}$ XX model on cylinders, from which we extract $-\ln(p_{\rm max})$,
the associated  $\ln L$ term, as well as an universal aspect-ratio dependent contribution of order $\mathcal{O}(1)$
that we compare to an analytical free-field calculation.
Sec.~\ref{sec:conc} is devoted to conclusions and discussion.

\begin{table}[htbp]
\begin{center}
\begin{tabular}{|lccc|}
\hline
Model  & $n$ &
\begin{tabular}{c}
$\ln(N)$ coef. \\
Ref.~\onlinecite{luitz_universal_2014}
\end{tabular}
& $\frac{N_\text{NG}}{4}\frac{n}{n-1}$ \\
\hline
\hline
Heisenberg  & &&   \\ 
$J_2=0$  & $\infty$ & 0.460(5) &  0.5 \\
$J_2=-5$ & $\infty$ & 0.58(2)  &  0.5 \\
\hline
$J_2=0$  & 2        & 1.0(2)  &  1 \\
$J_2=-5$ & 2        & 1.25(4) &  1 \\
\hline
&&&\\
$J_2=-5$ & 3        & 1.06(3) &  0.75 \\
\hline
&&&\\
$J_2=-5$ & 4        & 1.0(1) &   0.666 \\
\hline
\hline
\end{tabular}
\begin{tabular}{|lccc|}
\hline
Model  & $n$ &
\begin{tabular}{c}
$\ln(N)$ coef.\\
Ref.~\onlinecite{luitz_universal_2014}
\end{tabular}	
& $\frac{N_\text{NG}}{4}\frac{n}{n-1}$ \\
\hline\hline
XY	       &&&\\
$J_2=0$  & $\infty$ & 0.281(8) &  0.25 \\
$J_2=-1$ & $\infty$ & 0.282(3) &  0.25 \\
\hline
$J_2=0$  & 2        & 0.585(6) &  0.5 \\
$J_2=-1$ & 2        & 0.598(4) &  0.5 \\
\hline
$J_2=0$  & 3        & 0.44(2)  &  0.375 \\
$J_2=-1$ & 3        & 0.432(7) &  0.375 \\
\hline
$J_2=0$  & 4        & 0.35(8)  &  0.333 \\
$J_2=-1$ & 4        & 0.38(2)  &  0.333 \\
\hline
\hline
\end{tabular}

\end{center}
\caption{Subleading logarithmic terms in the SRE of the 2D Heisenberg and XY models, possibly with ferromagnetic second neighbor interaction $J_2$ (which strengthens the magnetic order).
$n$ is the Rényi (noted $q$ in Ref.~\onlinecite{luitz_universal_2014}).
The numerical values
obtained by Toulouse's group (supplementary material of~\onlinecite{luitz_universal_2014}) are
given in the third column. We selected  the best fit only for simplicity -- which  does not do justice to their
extensive and detailed data analysis.
The last column is the present theoretical prediction
(Eq.~\ref{eq:SnAll}), which combines the oscillators
(Eq.~\ref{eq:pmaxosc}) and TOS (degeneracy factor)
contributions (Eq.~\ref{eq:sntos}).  The
number $N_\text{NG}$ of Nambu-Goldstone modes is 2 for Heisenberg and 1
for XY.}  \label{tab:toulouse}
\end{table}

\section{Oscillator/spin-wave contributions}
\label{sec:osc}

\subsection{Massless free scalar field}

We first assume that the system is in a broken symmetry state, with a well-defined direction  
of the order parameter (say $x$).
At low energy the interactions between spin-waves are irrelevant and each mode
can be described by a free gapless scalar boson with a linear dispersion relation.
As a consequence we can  consider the case of a single mode ({\it i.e.} broken U(1)), and
the final result for the SRE  will simply have to be multiplied by the number of Nambu-Goldstone modes.

At each point $\mathbf r$ in space an angle $\phi_{\mathbf r}$  describes the local orientation
of the order parameter  with respect to its average direction.
At low energies and when coarse grained over sufficiently long distances, these deviations are small
and one can treat them as real numbers (instead of angles in $]-\pi,\pi]$), therefore neglecting the compactness of $\phi_{\mathbf r}$.
This leads to the Hamiltonian of a massless free scalar field:
\begin{equation}
 H = \frac{1}{2} \int d^2{\mathbf r} \left[ \chi_\perp  \Pi_{\mathbf r}^2 + \rho_s\left({\mathbf \nabla}\phi_{\mathbf r}\right)^2 \right]
 \label{eq:HFF}
\end{equation}
where $\rho_s$ is the stiffness,
$\chi_\perp=\frac{c^2}{\rho_s}$ is the transverse susceptibility, $c$ the spin-wave velocity, and
$\Pi_{\mathbf r}=\frac{\rho_s}{c^2}\dot \phi_{\mathbf r}$ is canonically conjugate to $\phi_{\mathbf r}$.
This is a collection of harmonic oscillators, one for each momentum $\mathbf k$:
\begin{equation}
 H = \frac{1}{2} \sum_{\mathbf k} \left[ \frac{c^2}{\rho_s} \Pi_{\mathbf k}^2 + \rho_s {\mathbf k}^2\left|\phi_{\mathbf k}\right|^2 \right].
 \label{eq:hff}
\end{equation}

\subsection{Configuration with the highest probability}

We start by considering the $n=\infty$ SRE, which amounts to evaluate
the probability of the ``most likely'' configuration.  As a warm up let us
first recall that the (normalized) ground-state wave function $\psi$ of
an harmonic oscillator with the Hamiltonian $H=\frac{1}{2m} p^2 +
\frac{1}{2}m\omega^2x^2$ is
\begin{equation}
 \psi(x)=\left(\frac{m\omega}{\pi}\right)^{1/4}\exp\left(-\frac{m\omega}{2}x^2\right).
\end{equation}
The probability density $p_{\rm max}$ to find the particle at its ``most likely'' location,
which is the square of the wave function at $x=0$, is the square of the normalization factor:
\begin{equation}
 p_{\rm max}=\left| \psi(0)\right|^2 = \left(\frac{m\omega}{\pi}\right)^{1/2}.
 \label{eq:ho1}
\end{equation}
Comparing this to Eq.~\ref{eq:hff}, the mode $\mathbf k$ of the free field has a mass $m_{\mathbf k}=\frac{\rho_s}{c^2}$ and frequency
$\omega_{\mathbf k}=c|\mathbf k|$. So, the probability $ p_{\rm max}(\mathbf k)$ for the mode $\mathbf k$ to be ``at the origin'' is:
\begin{equation}
 p_{\rm max}(\mathbf k) = \left(\frac{m_{\mathbf k}\omega_{\mathbf k}}{\pi}\right)^{1/2} = \left( \frac{\rho_s |\mathbf k|}{\pi c}\right)^{1/2}.
\end{equation}
We are interested in the probability density to observe $\phi_{\mathbf r}=0$ everywhere in space, so we impose $\phi_{\mathbf k}=0$ for all ${\mathbf k}$ and get:
\begin{eqnarray}
 p_{\rm max}^\text{osc} = \prod_{\mathbf k\ne 0} p_{\rm max}(\mathbf k) =\prod_{\mathbf k\ne 0} \left( \frac{\rho_s |\mathbf k|}{\pi c}\right)^{1/2}.
 \label{	}
\end{eqnarray}
Taking the logarithm we obtain:
\begin{eqnarray}
 - \ln\left( p_{\rm max}^\text{osc} \right) &=& -\frac{1}{2} \sum_{\mathbf k\ne 0} \ln \left( \frac{\rho_s}{\pi c}\right)
 - \frac{1}{4}\sum_{\mathbf k\ne 0} \ln  {\mathbf k}^2 .
 \label{eq:pmax}
\end{eqnarray}
The zero mode $\mathbf k=0$ is omitted since we assume that the system is in a broken-symmetry state.
Including the zero mode would, in a finite volume, ``delocalize'' the order parameter and  restore the rotation symmetry.
We will take later into account the rotational symmetry of the finite-size ground state by a correcting factor associated with the  ``degeneracy'' of the Anderson tower of states (TOS), see
Sec.~\ref{sec:TOS}.
The first sum in Eq.~\ref{eq:pmax} is simply a volume term ($\sim L^2$) but the universal contribution comes from the second
sum, which we analyze now.

\subsection{Determinant of Laplacian}
\label{ssec:Laplacian}

Since the $-{\mathbf k}^2$ are the eigenvalues of the Laplacian $\Delta$, the Eq.~\ref{eq:pmax} is a lattice regularization of $\ln \det' \Delta$, where  $\det'$
means that the zero eigenvalue is removed from the calculation of the determinant.

One can regularize the sum by using a periodic $L\times L$ lattice (torus), in which case
the universal terms in the $L\to\infty$ asymptotics can be extracted by means of an Euler-Maclaurin expansion. A possible way to regularize $\ln \det' \Delta$ is indeed to use the Brillouin zone of an $L\times L$ square lattice :
\begin{equation}
\Sigma(L) = \sum_{\mathbf k\ne 0} \ln \left( {\mathbf k}^2 \right)=  \sum'_{ n,m=-\frac{L}{2} \cdots \frac{L}{2}-1} \ln\left(k_n^2+k_m^2\right)
\end{equation}
where the discrete momenta are given by $ k_n=\frac{2\pi n}{L}$ and the zero-mode ($n=m=0$) is omitted.
Using twice the Euler-Maclaurin expansion at the trapezoid order gives:
\begin{eqnarray}
\Sigma(L) &=&  \left( \frac{1}{2}\pi -3-\ln\left( 2 \right) +2\ln\left( 2\pi\right)  \right) {L}^{2}\nonumber\\&& + \ln  \left( L^2\right) + \mathcal{O}(1).
\end{eqnarray}
While the term proportional to $L^2$ can be shown to depend on the regularization scheme, the $\ln  \left( L^2\right)$ is universal.

In fact, $ \det' \Delta$ is a quantity which has been studied extensively in the literature (see for instance Refs.~\onlinecite{kac_can_1966,duplantier_exact_1988}).
In particular, on a compact surface without boundary and with Euler characteristics $\chi$, one has:
\begin{equation}
  \ln \det{}' \Delta \simeq {\rm const.} L^2+\left(1-\frac{\chi}{6} \right) \ln\left({\rm L^2}\right).
  \label{eq:det_chi}
\end{equation}
This result is remarkable since the coefficient of the $\ln(L^2)$ term
is purely topological.  It can be derived using the heat-kernel method
and zeta regularization for instance~\cite{osgood_extremals_1988}.  An
explicit calculation, in cylinder geometry, is presented in Appendix
\ref{app:cyl_neu}.  We also note that, on a cylinder or on a torus, the
quantity $\ln \det' \Delta$ will also contain some finite aspect-ratio
dependent term, directly related to the one appearing in free boson
partition functions which are well studied in the context conformal
field theory \cite{di_francesco_conformal_1997}. 
The aspect-ratio dependent correction turns
out to be very important in the analysis of the numerical data presented
in Sec.~\ref{sec:dmrg}.

We therefore have:
\begin{equation}
 -\ln(p_{\rm max}^{\rm osc}) = {\rm const.} L^2 + \frac{1}{4} \left(\frac{\chi}{6}-1 \right) \ln  \left( L^2\right) + \mathcal{O}(1).
 \label{eq:topo}
\end{equation}
And, specializing to the torus ($\chi=0$):
\begin{equation}
 -\ln(p_{\rm max}^{\rm osc}) = {\rm const.} L^2  - \frac{1}{4} \ln  \left( L^2\right) + \mathcal{O}(1).
 \label{eq:pmaxosc}
\end{equation}
In the following, we are often interested only in the universal
logarithmic contribution and express, for example, Eq.~\ref{eq:pmaxosc}
as
\begin{equation}
  -\ln(p_{\rm max}^{\rm osc}) \sim  - \frac{1}{4} \ln{\left( L^2\right)} .
\end{equation}
If compared directly with the numerical QMC results for the $n=\infty$
SRE (Tab.~\ref{tab:toulouse}), the log coefficient
$-\frac{N_\text{NG}}{4}$ obtained above is clearly off, with a wrong
sign in particular. As we argue later, this is due to the fact that the
oscillator contribution provides only one part of the logarithmic
terms. The other part, discussed in Sec.~\ref{sec:TOS}, is due to the
fact that the ground state of a system of finite volume (as is the case
in the simulations) is rotationally invariant, contrary to the initial
assumption of a broken-symmetry state.  We note that this rotational
symmetry of finite systems also plays an important role concerning
logarithmic terms in the entanglement
entropy~\cite{metlitski_entanglement_2011,rademaker_tower_2015}.  Before dealing with this
important point (in the context of SRE), we discuss the $n$ dependence
of the oscillator contribution to the SRE.

\subsection{Finite Rényi index}
\label{ssec:finite_n}

So far we only considered one probability, $p_{\rm max}$, of observing the configuration with $\phi_{\mathbf r}=0$.
We will now discuss the $\ln(L)$ contribution to the finite-$n$ SRE.

Each probability $p_i$ (Eq.~\ref{eq:pi}) can be obtained in a path
integral formalism, by imposing the state $\ket{i}$ at $\tau=0$, the
plane corresponding to the imaginary time origin.  As already discussed
in the context of spin chains \cite{Oshikawa2010,stephan_phase_2011},
the quantity
\begin{equation}
Z_n=\sum_i p_i^n
\end{equation}
can be represented as an imaginary time path integral for the field theory
with $n$ replica fields $\phi^{(1)},\phi^{(2)},\ldots,\phi^{(n)}$.
Except at $\tau=0$, replica fields are decoupled, and each of
them is described by the same free boson field Lagrangian.
At $\tau=0$, we impose the ``gluing condition''
\begin{equation}
 \phi^{(1)} = \phi^{(2)} = \ldots = \phi^{(n)} .
\label{eq.gluing}
\end{equation}
This condition can be solved exactly, in a similar manner to
the analysis in 1 spatial dimension.
In fact, in general, we need to include possible boundary
perturbations, which turn out to be very important as we
will discuss below.

\subsubsection{Without boundary perturbations}

Keeping the caveat in mind,
first let us discuss what would be the SRE
in the absence of boundary perturbations.
In terms of the field theory, we can simply introduce the new
basis of the replica fields:
\begin{align}
 \Phi^{(0)} &= \frac{1}{\sqrt{n}} \sum_j \phi^{(j)}, \\
 \Phi^{(1)} &= \frac{1}{\sqrt{2}} \left( \phi^{(1)} - \phi^{(2)} \right),
\\
\vdots.
\end{align}
That is, $\Phi^{(0)}$ the ``center of mass'' field,
and the remaining $n-1$ fields
$\Phi^{(1)}, \ldots, \Phi^{(n-1)}$ are difference fields.
The gluing condition, Eq.~\ref{eq.gluing}, amounts to
imposing the Dirichlet boundary condition $\Phi^{(j)}=0$ for the
difference fields but leave the center-of-mass field $\Phi^{(0)}$
free~\cite{FradkinMoore-PRL2006,Hsu-EE-2dCQCP-PRB2009}.
It then follows that
\begin{equation}
 Z_n^{\rm osc} = \left( \frac{z_D}{z_0} \right)^{n-1},
\label{eq.Zn.nobp.z}
\end{equation}
where $z_D$ is the partition function for the single free boson
field with the Dirichlet boundary condition
imposed at $\tau=0$,
and $z_0$ is the partition function of the single free boson
without imposing any boundary condition at $\tau=0$.
Precisely speaking, the ``boundary'' $\tau=0$ is in the middle
of the entire system defined for $-\infty < \tau < \infty$
we consider. Nevertheless, it can still be regarded as a boundary
of $2n$-component boson field after a folding procedure~\cite{Oshikawa2010}.
Since imposing the Dirichlet boundary condition
is equivalent to freezing the fluctuation of the order parameter,
\begin{equation}
 p^{\rm osc}_{\rm max} = \frac{z_D}{z_0} .
\end{equation}
Thus we find
\begin{equation}
 Z_n^{\rm osc} \sim \left( p^{\rm osc}_{\rm max} \right)^{n-1},
\label{eq.Zn.osc.nobp}
\end{equation}
concerning the universal subleading contribution to $S_n$.
This would give
\begin{equation}
 S_n^{\rm osc}
 =\frac{1}{1-n}\ln Z_n^{\rm osc}
 \sim - \ln{p^{\rm osc}_{\rm max}}
 \sim - \frac{N_\text{NG}}{4}\ln{N} .
\label{eq.Sn.osc.nobp}
\end{equation}
which is actually the same as $S_{\infty}$.
For the free boson field theory in 1+1 dimensions,
the resolution of the gluing condition is in fact tricky
because of the subtlety in the compactification of the boson
field~\cite{Oshikawa2010}, leading to a correction
to the result as derived by the above argument.
However, in 2 spatial dimensions, the boson field can be
regarded as non-compact and the simple derivation as
given above stands correct.

The same result can be also derived without using replica trick,
following the analysis in 1 spatial dimension given in
Ref.~\cite{stephan_shannon_2009}.
Ignoring the possible boundary perturbations is equivalent to
consider the purely Gaussian wave function:
\begin{equation}
\psi(\left\{\phi_{\bf k}\right\})=\prod_{\bf k\ne 0} \left( \frac{\rho_s |\mathbf k|}{\pi c}\right)^{1/4} \exp\left(-\frac{\rho_s |\mathbf k|}{2c}\left|\phi_{\bf k}\right|^2\right).
\label{eq.Gaussian.wf}
\end{equation}
For such a state the calculation of $Z_n$ is just a Gaussian integration, and it can therefore be performed explicitly.
The result has a simple expression in terms of $p_{\rm max}^{\rm osc}$ (Eq.~\ref{eq:pmaxosc}):
\begin{equation}
 Z_n^{\rm Gauss}=\frac{\left(p_{{\rm max},\rho_s}^{\rm osc}\right)^n}{p_{{\rm max},n\rho_s}^{\rm osc}},
\end{equation}
where we have explicitly kept the dependence on the stiffness, and where the denominator
is evaluated at a modified value of the stiffness $\tilde \rho_s=n\rho_s$. 
For the massless oscillators discussed previously, the universal logarithm
in $\ln p_{{\rm max},\rho_s}^{\rm osc}$ is
actually independent of $\rho_s$ (see Eq.~\ref{eq:pmaxosc}).
Thus we find the same result as Eqs.~\ref{eq.Zn.osc.nobp}
and~\ref{eq.Sn.osc.nobp}.
This derivation has an advantage
that it is exact for an arbitrary real $n$ and does not rely on
the analytic continuation in $n$ which is usually required in a replica trick.
However, it should be still noted that it does rely on the assumption
of purely Gaussian wave function.
Even though such a Gaussian form correctly captures
the long-wavelength fluctuations of the order parameter, it does not
describe exactly the short-distance degrees of freedom
on the lattice. Neglecting the non-Gaussian terms in the wave function
corresponds to ignoring
the effects of possible boundary perturbations in the replica formulation.

\subsubsection{With the relevant boundary perturbation}

In the preceding analysis, we ignored the possible boundary perturbations,
which can be important.
In the replica field formulation, the replica fields are decoupled
in the bulk and each replica is described by the same Lagrangian density.
For the bulk,
we already know the asymptotically exact low-energy
effective theory, which corresponds to
the infrared fixed point of the renormalization group.
However, at the ``boundary'' ($\tau=0$) which is introduced by
taking the inner product with the basis states, the replica fields
are coupled and other boundary perturbations can arise.
In the presence of a relevant boundary perturbation, the
boundary condition is renormalized into a different one,
leading to a different SRE.
The general principle is that
all the boundary perturbations which are allowed by symmetries
would arise, unless they are eliminated by fine-tuning.
In SRE, because of the choice of the basis,
the U(1) symmetry is generally broken explicitly.

In fact, the change of boundary condition induced by
the boundary perturbation and 
the resulting ``phase transition'' in SRE were studied
in 1 spatial dimension~\cite{stephan_phase_2011}.
There, the leading boundary perturbation which is allowed
by the breaking of the U(1) symmetry and is consistent with
the compactification of the boson field is $\cos{\frac{\phi}{R}}$,
where $R$ is the compactification radius.
This implies, for the center-of-mass field, the boundary perturbation
$\cos{\frac{\Phi^{(0)}}{\sqrt{n}R}}$.
This is relevant for $n>n_c$.
Once relevant, it locks the center-of-mass field at the boundary,
giving rise to the Dirichlet boundary condition.

In contrast, in the 2D case discussed here, the boson field describes a small fluctuation on the
broken symmetry states, and thus it can be regarded as non-compact.
Therefore, we expect the boundary mass term $\sim \phi^2$
to appear, once the U(1) symmetry is broken.
The important difference from 1 dimension is that, the
boundary mass term is always relevant
(but see Sec.~\ref{sec:ndep}).
Its effect is still similar to 1 dimensional case,
locking the center-of-mass field at the boundary.
This results in the Dirichlet boundary condition on {\em all}
the $n$ replica fields.
Thus the partition function reads
\begin{equation}
 Z_n^{\rm osc} = \left( \frac{z_D}{z_0} \right)^{n},
\end{equation}
(compare with Eq.~\ref{eq.Zn.nobp.z} in the absence of the
boundary perturbation).
This leads to the universal logarithmic correction as
\begin{equation}
 S_n^{\rm osc} \sim - \frac{n}{n-1} \ln{p^{\rm osc}_{\rm max}}
  \sim 
 - \frac{N_\text{NG}}{4}\frac{n}{n-1}\ln(N) .
\label{eq.Sn.osc.bp}
\end{equation}

\section{Degeneracy factor}

\label{sec:TOS}

We have derived the universal oscillator contribution to SRE
in the previous section.
The final result for SRE, however, also requires a consideration
of the ground-state degeneracy due to the spontaneous symmetry breaking.

Let us briefly review the standard concept of tower of states (TOS)~\cite{anderson_approximate_1952,bernu_signature_1992,KomaTasaki-JSP1994,lhuillier_frustrated_2005}, which reconciles
the fact that the finite-size (antiferromagnetic) eigenstates are rotationally invariant while, in $D\geq2$,
the system can break the rotational symmetry in the infinite volume limit at $T=0$.

If a spin Hamiltonian $\cal H$ has a continuous rotation symmetry, say U(1) for simplicity,
the total angular momentum $S^z_\text{tot}=\sum_{\mathbf r} S^z_{\mathbf r}$ (generator of the rotations) is
a conserved quantity, and one can chose the eigenstates of $\cal H$ such that they are also eigenstates
of $S^z_\text{tot}$.
For an antiferromagnetic system, the finite-size ground state
has $S^z_\text{tot}=0$
and is thus rotationally invariant~\footnote{This was shown rigorously for an Heisenberg-like (or XXZ) model on a bipartite lattice (with the same number of sites on both sublattices):
the Lieb-Mattis theorem~\cite{lieb_ordering_1962}.}.

This may seem in contradiction with the possible spontaneous symmetry
breaking.
However, the symmetry of the finite-size ground state of course
does not rule out the possibility of the spontaneous symmetry
breaking.
Indeed, the spontaneous symmetry breaking is,
rigorously speaking, a concept which applies to the thermodynamic limit,
where ground states that break the symmetry must be
degenerate.

In order to realize some spontaneous symmetry breaking in the
thermodynamic limit, the finite-size spectrum must contain
low-energy eigenstates above the symmetric ground state.
Generic finite-size eigenstates of the Hamiltonian are also
eigenstates of $S^z_\text{tot}$, and thus each of them
does not break the symmetry.
The ``physical'' ground states in the thermodynamic limit,
which do break the symmetry,
correspond to superpositions of the finite-size low-energy eigenstates.
In the case of the spontaneous breaking of a continuous symmetry,
which is the focus of the present paper,
there must be an infinite number of such symmetry-breaking
physical ground states in the thermodynamic limit.
In order to produce these symmetry-breaking physical ground
states as superpositions, the number of the 
low-energy eigenstates in the finite-size spectrum
must grow as the system size is increased.
The set of these low-energy states (including the ground state) which reflect the spontaneous breaking
is commonly called Anderson TOS.

As discussed above, 
the finite-size counterpart of the symmetry-breaking ground states
(hereafter finite-size symmetry-breaking states for brevity)
are given by appropriate superpositions of the
finite-size eigenstates belonging to the Anderson TOS. 
This also implies that the symmetric finite-size ground state
is given by a superposition of the symmetry-breaking states.

It is helpful for understanding to map the spin system with
$S^z_\text{tot}$ conservation to
an interacting many-boson problem, by identifying
$S_\mathbf{r}^+$ with the creation operator $\psi^\dagger(\mathbf{r})$
and
$S_\mathbf{r}^-$ with the annihilation operator $\psi(\mathbf{r})$.
Then $S^z_\text{tot}$ corresponds to the total
number of particles, with a constant offset per site.
In a symmetry-breaking ground state
in the thermodynamic
limit $|\phi\rangle$,
$\psi(\mathbf{r})$ is thought to have a nonvanishing
expectation value, which can be regarded as an order parameter.
Specifically,
\begin{equation}
 \langle \phi | \psi(\mathbf{r}) | \phi \rangle
 = \sqrt{\rho_s} e^{i \phi},
\label{eq.SBSpsi}
\end{equation} 
where $\rho_s >0$ represents the superfluid density
and $\phi$ represents the phase of the condensate.
The symmetry-breaking ground state $|\phi \rangle$
is labeled by the phase $\phi$, a continuous parameter,
and thus is infinitely degenerate.

\begin{figure}
\includegraphics[width=8.5cm]{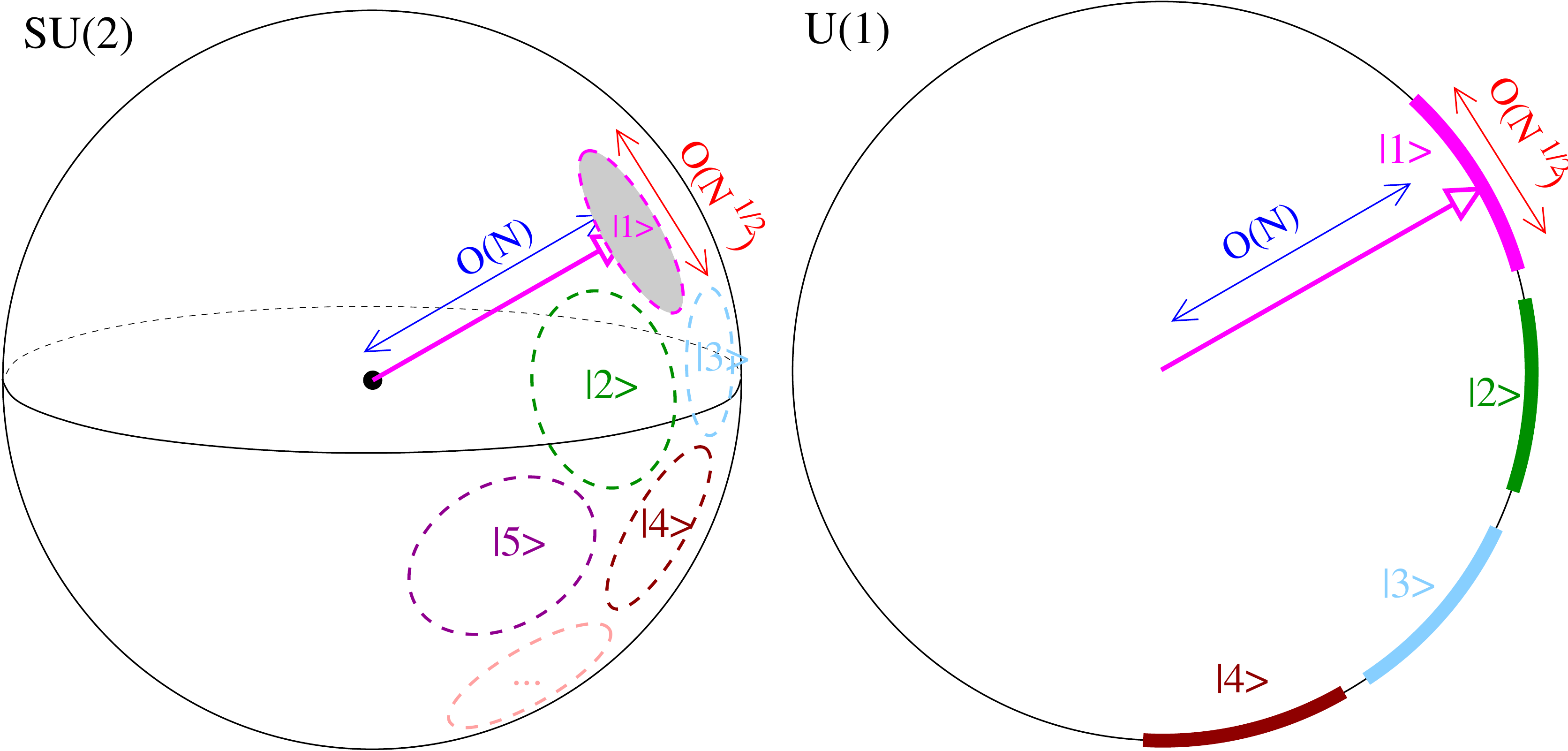}
\caption{Left: the order parameter fluctuations in the broken
symmetry states  $|i\rangle$ of a collinear SU(2) antiferromagnet are schematically represented
as patches on a sphere (the order parameter manifold).
Since the typical transverse fluctuations of the order parameter (sublattice magnetization) are $\sim N\Delta \phi=\mathcal{O}(N^{\frac{1}{2}})$ (Eq.~\ref{eq:Delta_phi}), each patch occupies an area $\sim \mathcal{O}(N)$.
From the fact that  the area of the sphere is $\sim N^2$, we get that the number $Q$ of non-overlapping patches scales as $\mathcal{O}(N)$.
Right: Case of a U(1) order parameter, where each broken symmetry state  $|i\rangle$ is represented by a (colored) arc on the circle.
Knowing that the transverse fluctuations are also $\sim N^{\frac{1}{2}}$, the phase space argument leads to  $Q \sim N^{\frac{1}{2}}$ independent states.
}
\label{fig:neel_vector}
\end{figure}

Now let us consider a finite-size system.
A finite-size symmetry-breaking state would also satisfy
Eq.~\ref{eq.SBSpsi}.
Such a state may be given as a coherent state satisfying
\begin{equation}
 \psi(\mathbf{r}) | \phi \rangle \sim 
\sqrt{\rho_s} e^{i \phi} | \phi \rangle .
\label{eq.coherent_phi}
\end{equation}
The expectation of the total number of particles
$N^p_\text{tot}$ in such a state is
\begin{equation}
 \langle N^p_\text{tot} \rangle = \sum_\mathbf{r} 
\langle \phi |
\psi^\dagger(\mathbf{r}) \psi(\mathbf{r})
| \phi \rangle = N \rho_s .
\end{equation} 
Likewise, we can also evaluate
\begin{eqnarray}
 \langle (N^p_\text{tot})^2 \rangle &=& \sum_{\mathbf{r} ,\mathbf{r}'}
\langle \phi |
\psi^\dagger(\mathbf{r}) \psi(\mathbf{r})
\psi^\dagger(\mathbf{r}') \psi(\mathbf{r}')
| \phi \rangle \nonumber \\ 
&=& (N \rho_s)^2 + N \rho_s .
\end{eqnarray} 
This implies a nonvanishing fluctuation
\begin{equation}
 \langle ( \Delta N^p_\text{tot} )^2 \rangle = N \rho_s .
\label{eq.DeltaN}
\end{equation}
The fluctuation of $N^p_\text{tot}$ (fluctuation of
$S^z_\text{tot}$ in the spin-system context)
is actually required by
the uncertainty relation
\begin{equation}
 {\Delta N^p_\text{tot} }{\Delta \phi} \gtrsim \frac{1}{2},
\label{eq.DeltaN_Deltaphi}
\end{equation}
which is a consequence of the non-commutativity
\begin{equation}
 [ N^p_\text{tot}, \phi ] \sim i .
\end{equation}
Eqs.~\ref{eq.DeltaN} and \ref{eq.DeltaN_Deltaphi}
implies that the finite-size symmetry-breaking state also
has an uncertainty in its phase:
\begin{equation}
 \Delta \phi = \mathcal{O}(\frac{1}{\sqrt{N}}).
 \label{eq:Delta_phi}
\end{equation} 
In other words, a symmetry-breaking
state ``occupies'' a finite patch on the circle representing all the possible order parameter directions,
as illustrated in Fig.~\ref{fig:neel_vector}~\footnote{Note that this scaling for the transverse fluctuations can also be  obtained using a linear spin-wave calculation.}.
This implies that two finite-size symmetry-breaking states
are distinguishable only if their phases differ by
more than $\Delta \phi = \mathcal{O}(\frac{1}{\sqrt{N}})$.

This can be also confirmed with the explicit construction of the
coherent state
\begin{equation}
 |\phi\rangle = e^{-\frac{\rho_s}{2}}
\exp{\left[ \frac{\sqrt{\rho_s}e^{i\phi}}{\sqrt{N}}
\sum_\mathbf{r}  \psi^\dagger(\mathbf{r}) \right]}
 | \text{vac} \rangle,
\label{eq.coherent_phi_exp}
\end{equation}
where $|\text{vac}\rangle$ is the vacuum with no boson present.
Using this expression, we find
\begin{align}
\left| \langle \phi | \phi' \rangle \right| &= 
\exp{\left[ - \rho_s N (  1 - \cos{|\phi-\phi'|} ) \right]}
\notag \\
& \sim 
\exp{\left[- \frac{1}{2} \rho_s N (\phi-\phi')^2 \right]},
\label{eq.coherent_overlap}
\end{align}
which is small when
$| \phi - \phi'| \gtrsim \mathcal{O}(\frac{1}{\sqrt{N}})$.
Therefore, in a finite-size system of $N$ sites,
there are $\mathcal{O}(\sqrt{N})$ linearly-independent symmetry-breaking states
in the case of the spontaneous breaking of a U(1) symmetry
\footnote{
From Eq.~\ref{eq.DeltaN} we see that the  states in the TOS
have a typical spin $S^z_{\rm tot}\sim \sqrt{N}$. Since the 
energies of the eigenstates in the TOS scale as $E\simeq \left(S^z_{\rm tot}\right)^2/N$~\cite{lhuillier_frustrated_2005} (the kinetic energy of
a quantum rotor with angular momentum $S^z_{\rm tot}$  and a moment of ``inertia''
proportional to the total number of spins),
 these states have a typical energy $E\sim \mathcal{O}(1)$
relative to the finite-size symmetric ground state. This corresponds
to a vanishing energy density in thermodynamic limit, as it should.
Finally, in the broken U(1) case the TOS is known to contain one eigenstate per value of $S^z_{\rm tot}$~\cite{lhuillier_frustrated_2005}.
The fact that the typical value of $S^z_{\rm tot}$ scales as $\sqrt{N}$ then implies a total number of state
is also $\mathcal{O}(\sqrt{N})$. This is consistent with what we found using the coherent state ansatz.
}.
It should be noted that 
the oscillator modes discussed in Sec.~\ref{sec:osc}
are not included in the above construction of the coherent states,
which are only used for counting the number of (almost)
independent symmetry-breaking ground states.
The final result on the SRE is obtained by combining
the counting of the symmetry-breaking ground states
and the contribution from the oscillator modes,
as it will done later in this paper.
We also note that, the simple coherent states discussed above
do not precisely represent physical symmetry-breaking ground states
in the presence of interactions (which is always the case
for quantum antiferromagnets)~\cite{Shimizu-Miyadera_PRL2000}.
Here those simple coherent states are used for simplicity,
as they should lead to the same number of independent
symmetry-breaking ground states.

The same argument, when applied to an SU(2) symmetry broken down to U(1) (collinear antiferromagnet) leads
to the conclusion that a low-energy  symmetry-breaking state occupies a 
solid angle $\delta\Omega  \sim 4\pi N^{-1}$ on the Bloch sphere representing the order parameter manifold (see Fig.~\ref{fig:neel_vector}).

In a more general situation we expect (phase space volume argument) the TOS dimension $Q$ to scale as $\sim N^\alpha$, 
with an exponent $\alpha$ which 
only depends on the number of Nambu-Goldstone modes:
\begin{equation}
 \alpha=N_\text{NG}/2.
 \label{eq:alpha}
\end{equation}
As discussed in the Introduction, throughout this paper
we consider systems with only type-I (type-A with linear dispersion)
Nambu-Goldstone modes, where the number of Goldstone modes
is equal to the number of broken symmetry generators.

\section{R\'enyi parameter ($n$) dependence of the SRE}
\label{sec:ndep}

Now that we have all the necessary ingredients, we shall
give the final results on the universal log correction to the SRE.
As we will discuss below, the results depends on
the R\'{e}nyi parameter $n$.

\subsection{$n>1$}

The choice of the $x$-basis explicitly breaks the U(1) symmetry 
of the XXZ model (or SU(2) symmetry of the Heisenberg model).
This  symmetry  breaking appears in the fact that, upon re-weighting the basis configurations according to $p_i\to p_i^n$,
the order parameter 
will preferentially be aligned with the $x$-direction.
This is obvious in the limit $n \to \infty$,
where the only configuration left is the
one with the largest probability, $p_{\rm max}$, and corresponds
to a perfectly ordered state with order parameter pointing in the $x$ direction.
While such a preferential direction is nontrivial for a finite $n$,
we expect it to hold for $n>1$ since, as we will discuss later,
the exact rotational symmetry is restored at $n=1$ only.
In terms of the field theory, such a preference is represented
by a boundary mass term.
Since such a mass term is always a relevant perturbation, we expect
Eq.~\ref{eq.Sn.osc.bp} to hold for general $n>1$.  However, as we have
mentioned earlier, Eq.~\ref{eq.Sn.osc.bp} contains only the oscillator
contributions.

What enters in the SRE is the probability $p_{\rm max}$, and, for the $x$-basis measurement of the antiferromagnetic XXZ model
we consider in this paper, we note $\ket{+}$ the associated spin configuration.
That is,\begin{equation}
 p_{\rm max} = \left| \langle \Psi | + \rangle \right|^2 .
\end{equation}
As we have discussed in Sec.~\ref{sec:TOS}, the symmetric finite-size
ground state $|\Psi\rangle$, with $S^z_\text{tot}=0$, is built as a
linear superposition of $Q\sim N^\alpha$ symmetry-breaking states noted
$\left\{ |1\rangle, |2\rangle, \cdots, |Q\rangle\right\}$.  The U(1)
case would correspond to $\alpha=\frac{1}{2}$ and SU(2)$\to$U(1) would
be $\alpha=1$ (see Fig.~\ref{fig:neel_vector}). We thus write:
\begin{equation}
 |\Psi\rangle =\frac{1}{\sqrt{Q}}\left( |1\rangle + |2\rangle + \cdots +
 |Q\rangle\right).
\label{eq:tos}
\end{equation}
As a consequence,
\begin{equation}
 p_{\rm max} = \left| \langle \Psi | + \rangle \right|^2 = \frac{1}{Q} \left| \sum_{i=1}^Q \langle i \ket{+} \right|^2.
\end{equation}
We can choose the states appearing in Eq.~\ref{eq:tos} so that only one, say $|1\rangle$, has an order parameter direction
which matches that  of the classical configuration $| + \rangle$.
We argue that $\sum_{i=1}^Q \langle i | + \rangle$ is dominated by the $i=1$ term, and that the others may be ignored
in the limit $N\to  \infty$, as
they are exponentially suppressed as a function of the system size $N$
relative to the dominant $i=1$ term.

On the other hand, since the state $\ket{1}$ is ``aligned'' with the
classical state $\ket{+}$, $\langle 1 | + \rangle$ will precisely have
the oscillator contribution as in Eq.~\ref{eq:pmaxosc}.
So, as far as the universal part is concerned, we may thus write
\begin{eqnarray}
 p_{\rm max} &\simeq &\frac{1}{Q}\left| \langle 1 | + \rangle \right|^2
 = \frac{1}{Q} p_{\rm max}^{\rm osc}
\end{eqnarray}
with $Q\sim N^{N_\text{NG }/2}$.  We finally get:
\begin{eqnarray}
 -\ln\left(p_{\rm max}\right) &\sim& -\ln\left(p_{\rm max}^{\rm
osc}\right) + \frac{1}{2}N_\text{NG}\ln(N) \label{eq:pmaxTOS}\\
&\sim& + \frac{1}{4}N_\text{NG}\ln(N) .
\end{eqnarray}

As already discussed in Sec.~\ref{ssec:finite_n},
we argue that, for $n>1$ where the boundary mass is relevant,
the universal contribution to the SRE is dominated by that of
$p_{\rm max}$ so that $S_n \sim \frac{n}{1-n}\ln(p_{\rm max})$.
Now $p_{\rm max}$ receives $\ln{N}$ contributions from
the gapless oscillator modes, as well
as from the degeneracy factor $Q$ discussed above. We may thus write
\begin{equation}
 S_{n>1}  \sim  \frac{n}{n-1}\left( \ln(Q)
 -\ln\left(p_{{\rm max}}^{\rm osc} \right)\right).
 \label{eq:sntos}
\end{equation}
Replacing $Q$ by $N^{N_\text{NG}/2}$ and $-\ln\left(p_{{\rm max}}^{\rm
osc} \right)$ by Eq.~\ref{eq:pmaxosc} we finally obtain:
\begin{equation}
 S_{n>1} \sim
 \frac{N_\text{NG}}{4}\frac{n}{n-1} \ln(N).  \label{eq:SnAll}
\end{equation}
In Tab.~\ref{tab:toulouse} the result above is compared to the QMC
results obtained by Luitz {\it et al.} (Toulouse group)~\cite{luitz_universal_2014}
at $n=2,3, 4$ and $\infty$.
The agreement is reasonable, and especially good for $n=\infty$,
although not perfect.
We also note that their
results for models {\em without} continuous symmetry breaking (gapped
phase of the XXZ model) indicate the absence of $\ln N$ correction,
which is of course consistent with the present analysis.
We stress that the error bars
given in Tab.~\ref{tab:toulouse} do not include the (significant)
variations when larger system sizes are included. For this reason we
believe that the numerical data are consistent with our predictions.
The apparently larger discrepancy between the theoretical prediction
and the numerical estimate is partly attributed to the smallness of
the boundary mass perturbation for smaller R\'{e}nyi parameter $n$.
In fact, as we will discuss in the next subsection (\ref{ssec:n=1}), the boundary
mass perturbation should vanish at $n=1$.
Thus, the crossover to the asymptotic behavior predicted theoretically
for $n>1$ would occur at larger lengthscale when $n$ is decreased
towards $1$.
We hope that further progress in numerical methods and theoretical
understanding of finite-size effects will improve the agreement.

An important support to the above reasoning is provided by the exact
result for the SRE of the Lieb-Mattis model~\cite{luitz_universal_2014}.
The latter has an SU(2)$\to$U(1) TOS (hence $\alpha=1$ and $Q\sim N$)
but no gapless spin-waves (hence no oscillator contribution to the
entropy). The ground state of this model was shown to have $S_{n>1}=
\frac{n}{n-1}\ln N+\mathcal{O}(1)$~\cite{luitz_universal_2014}, which
is in agreement with the first term (TOS contribution) in the r.h.s. of
Eq.~\ref{eq:sntos}.

\subsection{$n=1$}\label{ssec:n=1}

The case $n=1$ requires a special consideration.
When $n=1$, the boundary still retains the exact symmetry
of the Hamiltonian.
This can be seen because
\begin{equation}
 Z_1 = \sum_i p_i = \sum_i \langle \Psi | i\rangle \langle i | \Psi \rangle
  = \langle \Psi | \Psi \rangle = 1 .
\end{equation}
Namely, there is no particular boundary condition imposed at $\tau=0$;
it is rather a fictitious cut of the Euclidean space-time.

Even for $n=1$, the boundary mass could be added as a perturbation.
However, the exact symmetry discussed above implies that the
boundary mass perturbation is absent in the present problem
for $n=1$.
The absence of boundary mass corresponds to the quadratic action
and thus to the Gaussian wave function, Eq.~\ref{eq.Gaussian.wf}.
The Shannon entropy is then calculated
using the Gaussian wave function trick
as the $n \to 1$ limit of Eq.~\ref{eq.Sn.osc.nobp}.

The exact symmetry means that there is no preference given to
the direction of the order parameter.
Thus, all the $Q$ symmetry-breaking ground states contained
in the finite-size ground state (as in Eq.~\ref{eq:tos}) contribute to the universal
part of $p_{\rm max}$.
Therefore, unlike in the case of $n>1$, the $1/Q$
factor is missing, and the final result is given by
\begin{equation}
  S_{1} \sim - \ln{p^{\rm osc}_{\rm max}}
 \sim - \frac{N_\text{NG}}{4}\ln{N},
\label{eq.S1}
\end{equation}
where the universal logarithmic correction entirely comes from
the oscillator contribution.
The lack of the degeneracy factor can be indeed confirmed with the
exact result
\begin{equation}
  S_1 \sim 0
\end{equation}
for the Lieb-Mattis model~\cite{luitz_universal_2014},
in which there is no oscillator contribution.
The logarithmic correction in the Lieb-Mattis model comes only
from the degeneracy factor;
the fact that the $\ln N$ term precisely vanishes at $n=1$
implies that the degeneracy factor is also absent there,
reflecting the exact symmetry as discussed above.

\subsection{$n<1$}

The SRE is still well-defined for $n<1$.
In fact, it has been studied numerically 
for 1 spatial dimension, using exact numerical diagonalization~\cite{stephan_phase_2011}.
On the other hand, estimate of the SRE using Quantum
Monte Carlo simulation is more difficult for smaller $n$,
as contributions of smaller probabilities $p_i$ are more pronounced.
It is also difficult to perform simulations when $n$ is not an integer greater than one, since
this prevents the use of replica-based algorithms.
In fact, to our knowledge, no numerical data for the SRE at  $n<1$ is yet available in 2D.
Since analytical prediction of the SRE is also subtle for $n<1$,
in this paper we refrain from making a prediction in this regime and leave this question for future studies.

\section{Line subsystem}
 \label{sec:line}

So far we considered the configurations of the whole system, but it is also possible to consider the probabilities (and associated entropies)
of the configurations of a {\em subsystem}, noted $\Omega$.
For instance, the SRE of a segment in a critical spin chain was found to have some striking similarities with the entanglement entropy of that
segment~\cite{alcaraz_universal_2013,stephan_shannon_2014}.
In this section will specialize to the case where $\Omega$ is a line embedded in a 2D system.
In that case,
using  QMC and spin-wave calculations~\cite{luitz_universal_2015},
the {\em entanglement} entropy was recently shown to have some logarithmic correction.
We show here that the SRE possesses some very similar universal subleading term.

\subsection{Oscillators}

We first study the oscillator contribution to $p_{\rm max}^{\Omega}$, the probability of the most likely configuration of the region $\Omega$ (in the chosen basis).
For this, we consider the reduced density matrix of a subsystem in the framework of Eq.~\ref{eq:HFF}.
Since the Hamiltonian is Gaussian for the variables $\phi_r$, the reduced density matrix $\rho_{\Omega}$ is also Gaussian.
But to get the SRE entropies, and $p_{\rm max}^{\Omega}$ in particular, we do not need the full reduced density matrix but only its
diagonal elements. The latter, being again Gaussian,  must have the following form:
\begin{equation}
 \langle \phi |\rho_\Omega |\phi \rangle =\frac{1}{Z_\Omega} \exp\left(-\frac{1}{2} \sum_{\mathbf r  \mathbf r'\in \Omega} \phi_{\mathbf r}
  \left[\left(G_{|\Omega}\right)^{-1} \right]_{\mathbf r-\mathbf r'} \phi_{\mathbf r'} \right)
 \label{eq:pGp}
\end{equation}
where the state $|\phi \rangle$ has a fixed ``angle'' $\phi_{\mathbf r}$ at each site
and $(G_{|\Omega})_{\mathbf r,\mathbf r'}$ is the correlation function for two sites inside the region $\Omega$.
Using ${\rm Tr}\rho_\Omega=1$ and Gaussian integration, the normalization factor can be expressed using the determinant  of the correlation matrix $G_{|\Omega}$:
\begin{equation}
 Z_\Omega=\sqrt{\det\left[2\pi G_{|\Omega}\right]}.
\end{equation}
So, we already see that the probability $p_{\rm max}^{\Omega,\rm osc}$ to observe $\phi_{\mathbf r}=0$ everywhere in $\Omega$ is given by:
\begin{equation}
 p_{\rm max}^{\Omega,\rm osc} = 1/Z_\Omega = \left(\det\left[2\pi G_{|\Omega}\right]\right)^{-1/2}
\end{equation}
Or, in terms of the eigenvalues $g_{\mathbf k}$ of $G_{|\Omega}$:
\begin{equation}
 -\ln(p_{\rm max}^{\Omega,\rm osc})  = \frac{1}{2}\sum_{\mathbf k} \ln\left(2\pi g(\mathbf k)  \right).
 \label{eq:pl}
\end{equation}
Now we specialize the above calculation to the case where $\Omega$ is a line.
Due to the linear dispersion relation of the Goldstone mode, the long-distance behavior of the (transverse) correlation $G_{|\Omega}(r)$
is related to the (two-dimensional) Fourier transform of $1/k$, that is:
\begin{equation}
 G_{|\Omega}(r\to \infty)=G(r\to \infty) \sim 1/r .
\end{equation}
Now we transform this correlation back to real space, but restricting to the one-dimensional momentum $\mathbf k$ along the line.
We get:
\begin{equation}
 g({\mathbf  k}\to 0) \sim  -\ln(|{\mathbf  k}|).
\end{equation}
If we replace $g({\mathbf  k})$ by $ \sim -a\ln(|{\mathbf k}|)$ in Eq.~\ref{eq:pl} ($a>0$ is some non-universal factor)
and if we regularize the sum by taking a finite line with $L$ sites we obtain:
\begin{equation}
 -\ln(p_{\rm max}^{\rm line,\rm osc})  = \frac{1}{2}\sum_{\begin{array}{c}n=-L/2\\n\ne0\end{array}}^{L/2-1} \ln\left(-  2\pi a \ln\left(\frac{2\pi n}{L}\right)\right).
\end{equation}
This sum can be analyzed using an Euler-Maclaurin expansion. The dominant part turns out to be proportional to $L$, and the first subleading correction
turns out to be very slowly diverging:
\begin{equation}
 -\ln(p_{\rm max}^{\rm line,\rm osc})  = {\rm const.} L  - \ln\left(\ln(L)\right) + \mathcal{O}(1).
 \label{eq:loglog}
\end{equation}
In other words, there is {\it no} $\ln(L)$ term, contrary to the largest probability for the full system (compare with Eq.~\ref{eq:pmaxosc}). 

\subsection{Degeneracy factor}

The phase space argument of Sec.~\ref{sec:TOS} to treat the TOS contribution needs to be adapted for the
probability $p_{\rm max}^{\rm line}$ to observe an ordered configuration along a line.
Indeed, if we specify an ordered configuration $| \text{ord} \rangle$ only on a line, it involves $L$ sites only and the order parameter direction is fixed with a lower ``precision''.
Consequently we expect that {\em several} broken symmetry states
$|i\rangle$ (of the whole system) could have some significant
``overlap'' with $| \text{ord} \rangle$.

Let us examine the case of the U(1) symmetry breaking.
The symmetry-breaking ground state may be represented by
a coherent state.
The explicit expression~\ref{eq.coherent_phi_exp} can be also
written as
\begin{equation}
 |\phi\rangle = e^{-\frac{\rho_s}{2}}
\prod_\mathbf{r}
\{
\exp{\left( \frac{\sqrt{\rho_s}e^{i\phi}}{\sqrt{N}}
 \psi^\dagger(\mathbf{r}) \right)}
 | \text{vac} \rangle_\mathbf{r}
\}.
\end{equation}
This shows that the coherent state is a product state.

Fixing the spin configurations on the line amounts to
taking the partial trace of the ground-state density
matrix $|\Psi\rangle\langle \Psi|$
over the spin variables \emph{outside} the line,
and then projecting on the fixed spin
configuration on the line.
As we argued earlier, 
the finite-size symmetric ground state $|\Psi\rangle$
may be written as a superposition of almost independent
symmetry-breaking (coherent) states as in Eq.~\ref{eq:tos}.
We thus first write the reduced density matrix of the line:
\begin{equation}
 \rho_{\rm line} = \frac{1}{Q}\sum_{i,j=1}^Q {\rm Tr}_{\rm \overline{\rm line}}\left(|i\rangle\langle j|\right)
\end{equation}
where the trace is performed over the degrees of freedom lying outside the line.
Because the exterior of the line is a large subsystem ($\sim N$ sites) it seems clear that
no state $|e\rangle$ outside the line
can achieve a significant overlap simultaneously with $|i\rangle$ and $|j\rangle$ if $i\ne j$.
Furthermore, since the coherent state is a product state,
the partial trace can be carried out to obtain
\begin{eqnarray}
\rho_{\rm line}
  &\sim & \frac{1}{Q}
\sum_{j=1}^Q 
\left|\phi=\frac{2\pi j}{Q}\right\rangle_\text{line} \mbox{\scriptsize{ }}
\leftidx{_\text{line}}{\left\langle \phi=\frac{2\pi j}{Q}\right|},
\end{eqnarray}
where $|\phi\rangle_\text{line}$ is a coherent state defined on the line.
However, an evaluation of the overlap between the coherent states
on the line similar to Eq.~\ref{eq.coherent_overlap}
reveals that they are independent only if the
angle parameters differ by $O(1/\sqrt{L})$ or more.
Thus, in terms of the (almost) independent coherent states
on the line,  
\begin{equation}
\rho_{\rm line}
  \sim \frac{1}{\tilde{Q}}
\sum_{j=1}^{\tilde{Q}} 
\left|\phi=\frac{2\pi j}{\tilde{Q}}\right\rangle_\text{line} \mbox{\scriptsize{ }}
\leftidx{_\text{line}}{\left\langle \phi=\frac{2\pi j}{\tilde{Q}}\right|},
\end{equation}
where $\tilde{Q}=\mathcal{O}(\sqrt{L})$ and the
overall factor is determined by the condition
$\Tr_\text{line}\rho_{\rm line} = 1$.

Thus we find
\begin{equation}
 p_{\rm max}^{\rm line} \simeq  \frac{1}{\tilde{Q}} p_{\rm max}^{\rm line,osc}, 
\end{equation}
or equivalently
\begin{eqnarray}
 -\ln{\left(p_{\rm max}^{\rm line}\right)}
 &\simeq&  -\ln{\left(p_{\rm max}^{\rm line,osc}\right)} 
 + \frac{1}{2} \ln{\left(L\right)}
 \label{eq:ltos_U1}
\end{eqnarray}

Similar arguments can be constructed for the SU(2) case, leading to
a $\ln\left(L\right)$ term.
More generally, we may conjecture that the result only depends on the number of Goldstone modes:
\begin{eqnarray}
 -\ln\left(p_{\rm max}^{\rm line}\right) &\simeq&  -\ln\left(p_{\rm max}^{\rm line,osc}\right)  + \frac{N_{\rm NG}}{2}\ln\left(L\right).
 \label{eq:ltosNG}
\end{eqnarray}

\subsection{Final result and comparison with the numerics}
As just done for the whole system, we can combine the oscillator contribution
({\it i.e.} no $\ln L$ term, see Eq.~\ref{eq:loglog}), the TOS contribution (Eq.~\ref{eq:ltosNG})
and the argument of Sec.~\ref{sec:ndep}
to get the $n$ dependence.
The final result for the SRE is
\begin{equation}
 S_{n>1}^{\rm line} \sim  \frac{N_\text{NG}}{2}\frac{n}{n-1} \ln(L).
 \label{eq:SnLine}
\end{equation}
For $n=\infty$,   Luitz {\it et al.}~\cite{luitz_shannon-renyi_2014} found the coefficient of $\ln(L)$  to be $\gtrsim 0.7$ for a system
with $N_\text{NG}=2$ (to be compared to $1$ from
the formula above).
In a more recent work~\cite{luitz_quantum_2016} the  QMC calculations
were pushed up to $L=40$ for $n=2,3,4$, up to $L=128$ for $n=\infty$ and up to $L=30$ for non-integer values $n$.
In all cases the QMC results are in good agreement with Eq.~\ref{eq:SnLine}.

\section{$p_{\rm max}$ for the 2D spin-$\frac{1}{2}$ XY model on the square lattice}
\label{sec:dmrg}

\begin{figure}[h]
 \includegraphics[angle=-90,width=9cm]{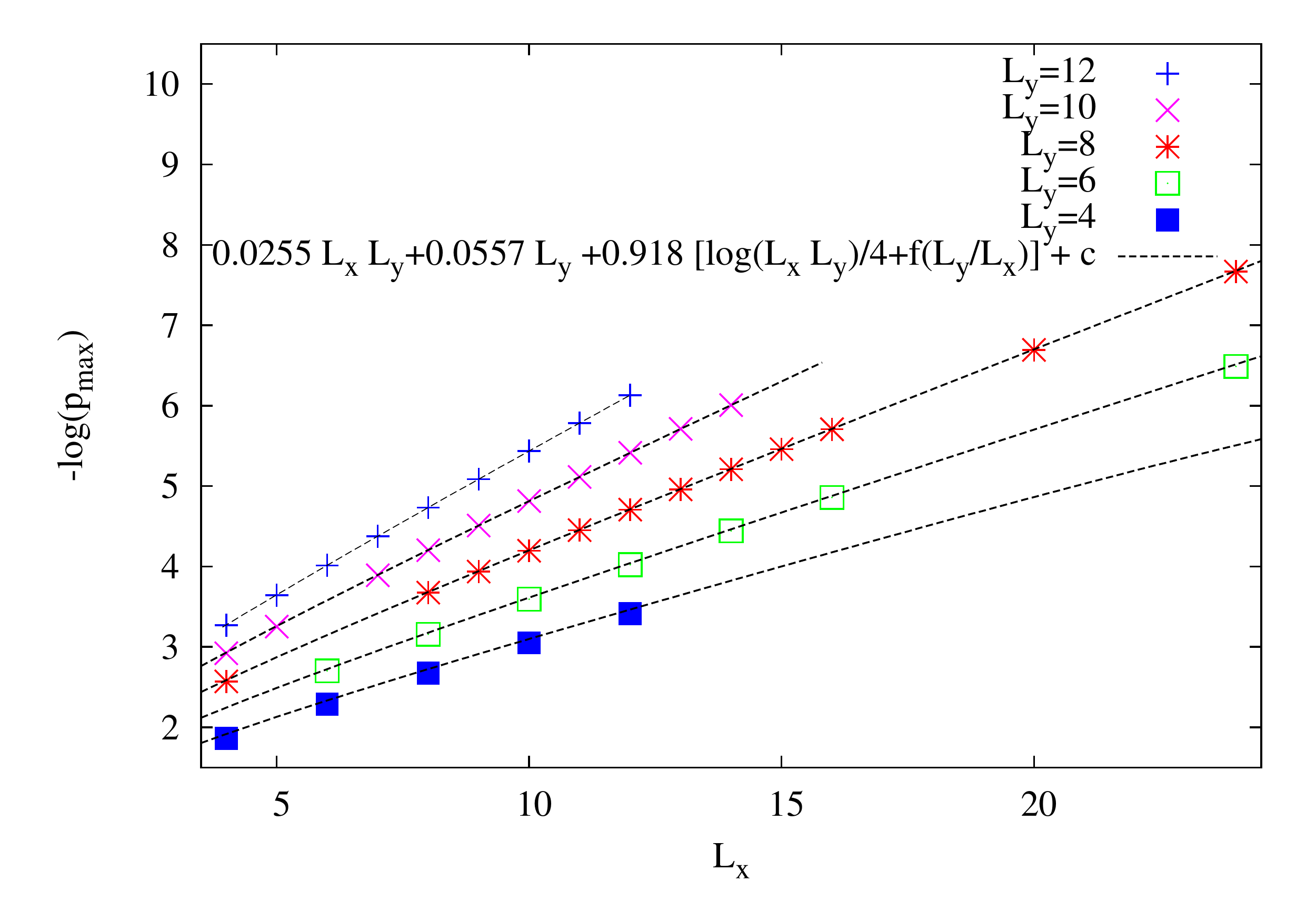}
 \caption{DMRG results for $p_{\rm max}$ in the 2D XY ferromagnet (data given in Tab.~\ref{tab:pmax_dmrg}).
 The fitting function (fit restricted to the data points with $L_{x,y}\geq10$, see text) is shown with dashed lines. The prefactor of the logarithm term, here $0.918/4$ is
 in good agreement with the theoretical prediction for a single Nambu-Goldstone mode ($1/4$).
 The aspect ratio-dependent term $f_N(L_y/L_x)$, defined in Eq.~\ref{eq:fN}, contains no free parameter.
 }
 \label{fig:pmax_dmrg}
\end{figure}

In order to provide some additional check for our predictions concerning $p_{\rm max}$, we consider the ferromagnetic XY model on the square lattice:
\begin{equation}
 H=-\sum_{\langle i,j\rangle} \left( S^x_i S^x_j + S^y_i S^y_j \right),
\end{equation}
which spontaneously breaks the U(1) symmetry in the thermodynamic limit ($N_{\rm NG}=1$).
The ground state $\left|\psi \right\rangle$
in the $S^z_{\rm tot}=0$ sector
was obtained numerically using 
2D DMRG \cite{stoudenmire_studying_2012} (using the C++ iTensor library \cite{itensor}) on cylinders of length $L_x$ and circumference $L_y$,
up to $L_y=12$. The probability $p_{\rm max}$ is defined by projection onto the state where
all spins point in the (say) $x$ direction:
\begin{equation}
 p_{\rm max}= | \left \langle |\psi | \rightarrow \cdots \rightarrow \right\rangle |^2.
\end{equation}
Once $\left|\psi \right\rangle$ is in a matrix-product form (as produced by the DMRG algorithm), $p_{\rm max}$ is easily obtained
by computing the scalar product with the ferromagnetic configuration above (a product state).
The numerical results are given in Tab.~\ref{tab:pmax_dmrg} and plotted in Fig.~\ref{fig:pmax_dmrg}. The matrix dimensions (up to $\chi=6000$) were chosen to insure that the maximum truncation error
stays below $10^{-7}$ for $L_y< 12$ and below $5.10^{-7}$ for $L_y= 12$. This insures a precision of at least four digits
on $p_{\rm max}$ for the largest systems.

We now discuss the theoretical prediction for $p_{\rm max}$ in the
cylinder geometry.  First, the TOS contribution is expected to be
independent of the geometry and should therefore be (single
Nambu-Goldstone mode, see Eq.~\ref{eq:pmaxTOS}):
\begin{equation}
 -\ln(p_{\rm max}^{\rm TOS}) =\frac{1}{2} \ln N,
\end{equation}
where $N=L_xL_y$. As for the torus, the oscillator contribution to $-\ln(p_{\rm max})$
has a non-universal ${\rm const.} N$ term, and some universal part
related to the determinant of the Laplacian:
\begin{equation}
 -\ln(p_{\rm max}^{\rm osc}) \sim 
 -\frac{1}{4}\ln \det{}' \Delta.  \label{eq:pmax_det}
\end{equation}
The leading universal part is a $\ln(N)$ term related to the Euler
characteristics $\chi$ (see Eq.~\ref{eq:det_chi}). From the fact that
$\chi=0$ on cylinder, we have $\ln \det' \Delta\sim \ln(N)$.  Adding
the TOS contribution one gets:
\begin{equation}
 -\ln(p_{\rm max}^{\rm osc+TOS,cyl.}) \sim
  \frac{1}{4}\ln\left(N\right).
\end{equation}
In practice, the accessible system sizes are not large enough to extract
from $-\ln(p_{\rm max})$
the coefficient of the $\ln(N)$ directly and reliably.  To analyze the
finite-size data of Tab.~\ref{tab:pmax_dmrg}, it is therefore
interesting and useful to look also for the next subleading term in
$\ln \det' \Delta$.  The latter is finite in the thermodynamic limit,
and it depends in some universal manner on the aspect ratio $r=L_y/L_x$
of the cylinder. Such terms are well known in the context of partition
functions in 2D conformal field theory, since the determinant of the
Laplacian is related to the (non-compact) free-boson partition function
(see Eq. 10.16 in \cite{di_francesco_conformal_1997}) :
\begin{equation}
 Z_{\rm free\; boson}=\sqrt{\frac{A}{\det{}' \Delta}},
\end{equation}
where $A$ is the area (here $A=N=L_xL_y$).

In our case we have
a cylinder with free spins at the boundaries. This translates to some free boundary conditions (BC) for the oscillators, and such
conditions are expected to flow (in the renormalization group sense) to some Neumann BC for the free field.
So, we need to compute the determinant of the Laplacian on a cylinder with Neumann BC.
This quantity can be computed using zeta-regularization \cite{osgood_extremals_1988},
as detailed in Appendix \ref{app:cyl_neu}. The result is:
\begin{equation}
  \det{}'\Delta = A r \left|\eta\left(\frac{ir}{2}\right)\right|^2
  \label{eq:det_cyl}
\end{equation}
where $\eta$ is the Dedekind $\eta$-function. Plugging this result in Eq.~\ref{eq:pmax_det}  
gives
\begin{equation}
 -\ln(p_{\rm max}^{\rm osc,cyl.}) \sim -\frac{1}{4}\ln\left(N\right) -\frac{1}{2}\ln\left[\sqrt{r}  \eta\left(\frac{ir}{2}\right)\right].
\end{equation}
We finally add the TOS contribution to get:
\begin{eqnarray}
 -\ln(p_{\rm max}^{\rm osc+TOS,cyl.}) &\sim&\frac{1}{4}\ln\left(N\right)\nonumber\\
 &&-\frac{1}{2}\ln\left[\sqrt{\frac{L_y}{L_x}} \eta\left(\frac{iL_y}{2L_x}\right)\right].
 \label{eq:pmax_cyl}
\end{eqnarray}
So, we analyzed the data with the following fitting function:
\begin{eqnarray}
 -\ln(p_{\rm max})&\simeq&a L_x L_y + bL_y +c \label{eq:fit}\\
  &&+ d\left[\ln(L_x L_y)/4 +f_N\left(\frac{L_y}{L_x}\right)  \right] \nonumber \\
  {\rm with}\,\,\,f_N(r)&=&-\frac{1}{2}\ln\left[\sqrt{r} \eta\left(\frac{ir}{2}\right)\right] \label{eq:fN},
\end{eqnarray}
and $a$, $b$, $c$, and $d$ are four free parameters. From our theoretical analysis (Eq.~\ref{eq:pmax_cyl}), $d$
corresponds to the number of Nambu-Goldstone mode(s)
and should be close to 1.
The result of the fits is shown in Fig.~\ref{fig:pmax_dmrg}. The dashed lines
represent a fit to the data points with $L_x,L_y\geq 10$, and gives $d=0.918$.
We note that, although only the largest system sizes were used in the fit, the function defined in Eq.~\ref{eq:fit}
goes through all the data points with a relatively good accuracy, including the small systems.
We also mention that the parameter $d$ we have obtained is relatively stable:
we find $d\simeq 0.906$ if we restrict the fit to the cylinders with $L_{x,y}\geq 8$,
$d\simeq 0.899$ if we restrict to $L_{x,y}\geq 6$,
and $d\simeq 0.915$ if we used all the data (including $L_{x,y}$ as small as 4).
Although we have not performed a precise analysis of the error bar, our experience with varying the number of data points included in the fit
indicates that the data we described well by $N_{\rm GN}\simeq d=0.9(1)$.

\begin{figure}[h]
 \includegraphics[angle=-90,width=9cm]{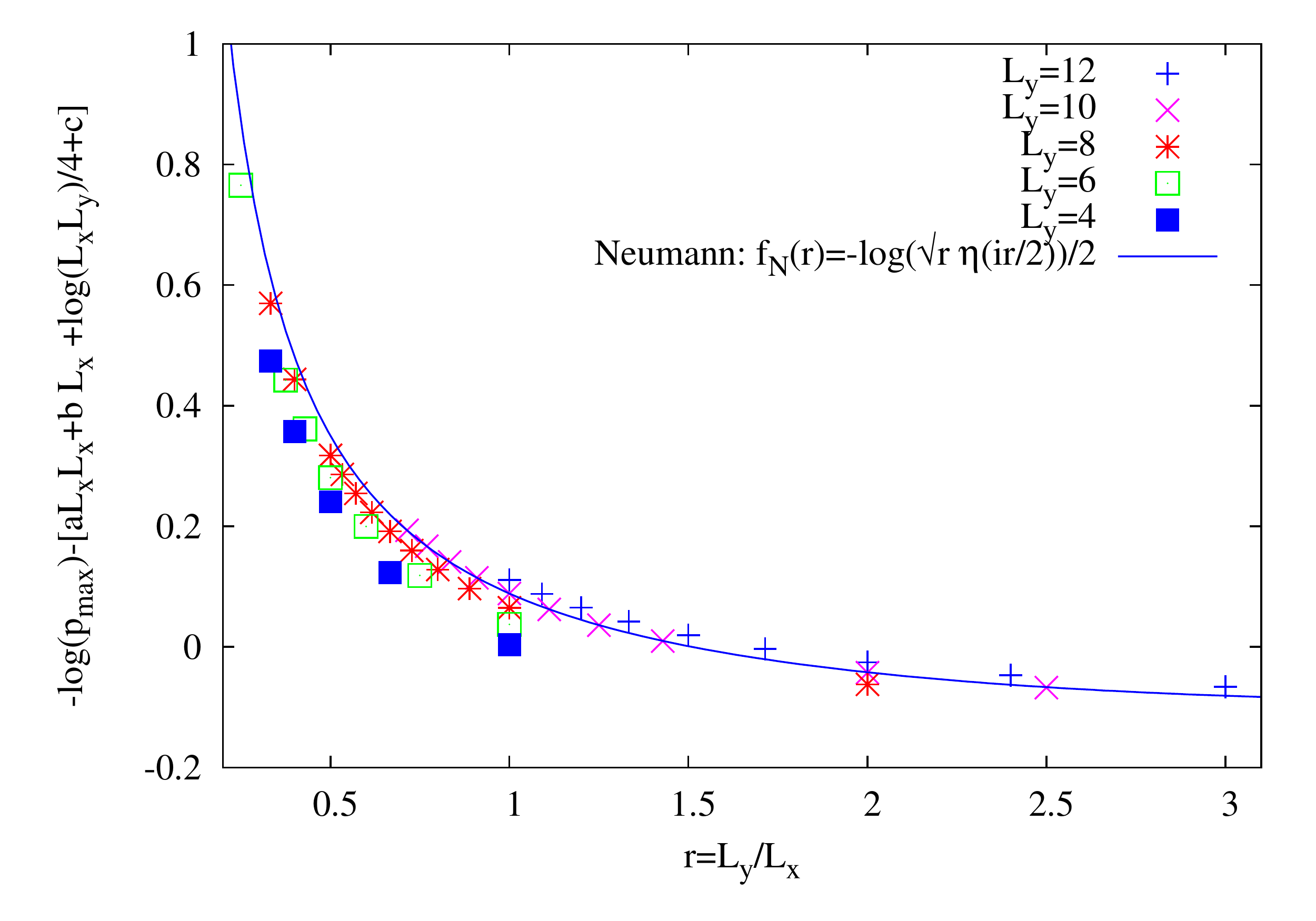}
  \caption{
 The DMRG results (Tab.~\ref{tab:pmax_dmrg} and Fig.~\ref{fig:pmax_dmrg})
 are fitted to $g(L_x,L_y)=aL_xL_y + bL_y + c+\ln\left(L_xL_y\right)/4+f_N(L_y/L_x)$ (three adjustable parameters: $a$, $b$, and $c$).
 As for Fig.~\ref{fig:pmax_dmrg}, the fit was restricted to the data points with $L_{x,y}\geq10$.
 The plot represents the difference between the  data and $aL_xL_y + bL_y + c+\ln\left(L_xL_y\right)/4$, as a function of the cylinder aspect-ratio $r=L_y/L_x$. This
 difference is well described by the aspect-ratio dependent term $f_N(L_y/L_x)$ (Eq.~\ref{eq:fN}) that is predicted for Neumann boundary conditions at the edges of the cylinder.
 }
 \label{fig:aspect_ratio}
\end{figure}

To check further the validity of this analysis, we have fitted the data 
by the function above, but imposing $d=1$. This leaves three  free parameters: the area coefficient $a$, the
linear coefficient $b$, and a constant $c$. We have plotted in Fig.~\ref{fig:aspect_ratio} the difference between the numerical data
and $a L_x L_y + bL_y +\ln(L_x L_y)/4 + c$ . These variations, plotted as a function of the aspect ratio $r=L_y/L_x$, appear to be very well described 
by $f_N(r)$, as expected if there is an underlying free boson system with Neumann boundary conditions.
The agreement between the data and $f_N$ is quite good, given the fact the plot contains only three adjustable parameters.

\begin{table}[!htbp]
\begin{tabular}{|cc|ll|ccl|}
\hline
 $L_x$ 	& $L_y$ & $E$ 			& $-\ln(p_{\rm max})$ 	&$\chi$& sweeps& error \\
 \hline
 4 	& 4	& -8.0167741	&1.864172	&256	& 17 	& 0		\\
 6 	& 4 	&-12.4272461	&2.285488	&800 	& 15 	& 5.96e-14	\\
 8 	& 4	&-16.8429370	&2.675768	&800	& 16 	& 2.44e-13	\\
10	& 4	&-21.2608268	&3.049265	&800	& 16	& 7.944e-13	\\
12	& 4	&-25.6798523	&3.412229	&100	& 20	& 3.753e-13	\\
 \hline 
 6 	& 6 	&-18.4620013 	&2.701232	&1000 	& 33 	& 1.17e-09	\\
 8	& 6	&-25.0549930	&3.155927	&1000	& 29	& 1.83e-09	\\
10	& 6	&-31.6501154	&3.594168	&1000	& 26 	& 2.85e-09	\\
12	& 6	&-38.2463365	&4.021994	&2000	& 41	& 1.65e-10	\\
14	& 6	&-44.8431957	&4.442731	&2600	& 30	& 6.43e-11	\\
16	& 6	&-51.4404602	&4.858276	&2600	& 36	& 9.10e-11	\\
24	& 6	&-77.8315765	&6.488096	&2600	& 31	& 2.32e-10	\\
\hline 
4	& 8	&-15.7729479	&2.570953	&3000	&22	&5.33e-09	\\
8	& 8	&-33.3327539	&3.674920	&3000	&50	& 4.47e-09 	\\
9	& 8	&-37.7249437	&3.936932	&3000	&41	& 5.07e-09 	\\	
10	& 8	&-42.1174937	&4.195792	&3000	&33	& 5.91e-09 	\\
11	& 8	&-46.5103127	&4.452070	&4000	&50	& 2.04e-09	\\
12	& 8	&-50.9033297	&4.706283	&4000	&50	& 2.36e-09 	\\
13	& 8	&-55.2965038	&4.958716	&4500	&40	& 1.73e-09 	\\
14	& 8	&-59.6898005	&5.209670	&4500	&50	& 1.98e-09 	\\
15	& 8	&-64.0831974	&5.459323	&4800	&50	& 1.73e-09	\\
16	& 8	&-68.4766749	&5.707869	&4800	&50	& 1.93e-09	\\
20	& 8	&-86.0511590	&6.693276	&4800	&50	& 3.00e-09	\\
24	& 8	&-103.626222	&7.668465	&4800	&50	& 3.92e-09	\\
\hline
4	& 10	&-19.6833386	&2.921201	&4000	&38	&9.21e-08	\\
5	& 10	&-25.1667352	&3.252921	&4000	&46	& 6.61e-08	\\
7	& 10	&-36.1399647	&3.891643	&4000	&50	& 6.70e-08	\\
8	& 10	&-41.6279980	&4.202464	&4000	&49	& 6.83e-08	\\
9	& 10	&-47.1165448	&4.509336	&4000	&50	& 7.40e-08	\\
10	& 10	&-52.60545201	&4.813080	&4000	&50	& 8.13e-08	\\
11	& 10	&-58.0946827	&5.114146	&5000	&45	& 4.60e-08	\\
12	& 10	&-63.5840956	&5.413127	&6000	&40	& 2.90e-08	\\
13	& 10	&-69.0736369	&5.710448	&6000	&45	& 3.12e-08	\\
14	& 10	&-74.5633010	&6.006292	&6000	&46	& 3.40e-08	\\
\hline
4	& 12	&-23.5988117	&3.268168	&6000	&45	& 3.64e-07	\\
5	& 12	&-30.1784875	&3.644283	&6000	&50	& 2.88e-07	\\
6	& 12	&-36.7611498	&4.012780	&6000	&50	& 2.81e-07	\\
7	& 12	&-43.3450519	&4.374877	&5000	&50	& 4.34e-07	\\
8	& 12	&-49.9299175	&4.732248	&5000	&42	& 4.36e-07	\\
9	& 12	&-56.5153082	&5.085815	&5000	&50	& 4.56e-07	\\
10	& 12	&-63.1010587	&5.436351	&5000	&50	& 4.85e-07	\\
11	& 12	&-69.6874332	&5.784312	&6000	&37	& 3.28e-07	\\
12	& 12	&-76.2736890	&6.130302	&6000	&34	& 3.47e-07	\\

\hline
\end{tabular}
 \caption{DMRG results for $p_{\rm max}$ in the 2D XY ferromagnet.
 $L_x$ is the length of the cylinder, and $L_y$ is the perimeter. $E$ is the ground-state energy.
 Due to the area-law scaling of the entanglement entropy, $ \chi$
 should grow exponentially with $L_y$ to insure an accurate description of the wave function. The last column provides the largest truncation error measured
 during the last DMRG sweep.}
 \label{tab:pmax_dmrg}
\end{table}

\section{Conclusion}
\label{sec:conc}

We have shown theoretically that a spontaneously broken continuous
symmetry leads to some universal logarithmic contribution to the
Shannon-Rényi entropies.
By connecting the Shannon-R\'{e}nyi entropy
due to Nambu-Goldstone mode fluctuations with the determinant of the
Laplacian, we also showed that the logarithmic contribution to the
Shannon-Rényi entropy is topological and depends on the Euler-Poincaré
characteristics of the two-dimensional system.
As the ground state of a finite-size system is symmetric while the
choice of the basis selects a particular symmetry-broken state,
there is an additional logarithmic contribution to the Shannon-R\'{e}nyi
entropy corresponding to the ground-state degeneracy.
Combining the two contributions together, the universal logarithmic
term in the Shannon-R\'{e}nyi entropy we find is in a good
agreement with the numerical result obtained by the Toulouse group. 
We have also extended our analysis to the Shannon-R\'{e}nyi entropy
defined with respect to a line subsystem.

The situation turns out to be remarkably similar to the logarithms found
in the {\it entanglement} entropy of a subsystem, where, also, the
zero-point motion of the oscillator modes and the rotational symmetry of
the finite-size ground state had to be
included~\cite{metlitski_entanglement_2011}.
This suggests that there is a deep connection between the
entanglement and Shannon-R\'{e}nyi entropies, despite the obvious
differences such as partition dependence of the former and
the basis dependence of the latter. 
In fact, the Shannon-R\'{e}nyi entropy has been also discussed
in the context of the entanglement entropy in systems at
conformal critical points.
There, the entanglement entropy in a certain class of
wave functions in $D+1$ spatial dimensions is
mapped to the Shannon-R\'{e}nyi entropy
in a $D$-dimensional system~\cite{stephan_shannon_2009}.
On the other hand, given the similarity of the present analysis to
that in Ref.~\cite{metlitski_entanglement_2011},
there might be a direct connection between the two different
entropies in the same $D$-dimensional system.
Elucidation of such a connection would be useful to advance
further our understanding on both entropies.

As in the case of the entanglement entropy, our hope is
that the Shannon-R\'{e}nyi entropy will be useful
as a diagnostic tool to characterize and classify
quantum phases, in particular those beyond the traditional
classification scheme.
Our efforts in the present paper is limited to the conventional
phases with a spontaneous broken continuous symmetry, which
are already understood very well.
Nevertheless, the fact that we can extract the number of the
Nambu-Goldstone modes from the scaling of the Shannon-R\'{e}nyi
entropy suggests that this could be a novel tool as useful
as the entanglement entropy.
In order to extend the application of the Shannon-R\'{e}nyi
entropy to less conventional phases, it would be important
to develop a new numerical scheme as well as analytical methods,
since many of interesting phases arise in the presence
of frustration which often makes quantum Monte Carlo simulations
difficult.

A relatively straightforward extension of the present work
would be to study the effects of sharp corners in the system geometry,
which would also contribute to the logarithmic divergence
of the Shannon-R\'{e}nyi entropy.
Checking theses property numerically would provide valuable tests for
the arguments presented here.

The symmetry argument presented in Sec.~\ref{sec:ndep}
suggests that the value $n=1$ of the Rényi index corresponds to
a phase transition point, with a quite different predicted
behavior of the SRE (Eq.~\ref{eq.S1}).
Numerical verification of our prediction would be an interesting
problem.
Further elucidation of this phase transition, and exploration
into the $n<1$ phase also seems an interesting direction of research,
both from the analytical and numerical point of views.

\section*{Acknowledgments}
We wish to thank Fabien Alet, David Luitz and Nicolas Laflorencie for
many useful discussions and sharing some of their unpublished data.  We
also thank Fabien Alet for numerous insightful comments on the
manuscript.  M.~O. is also grateful to Haruki Watanabe on useful
discussions on related subjects, and to IRSAMC Toulouse and CEA Saclay
for the hospitality during his visits where parts of this work were
carried out.  V.~P. is grateful to Stéphane Nonnenmacher for introducing
him to Ref.~\onlinecite{osgood_extremals_1988}.  G.~M. is supported by a
JCJC grant of the Agence Nationale pour la Recherche (Project
No. ANR-12-JS04-0010-01).  M.~O. is supported in part by JSPS KAKENHI
Grant Nos. 25400392 and 16K05469, and US National Science Foundation
under Grant No. NSF PHY1125915 through Kavli Institute for Theoretical
Physics, UC Santa Barbara.

\appendix

\section{Laplacian determinant on cylinder with Neumann B.C.}
\label{app:cyl_neu}

We consider a cylinder of length $L_x$ and circumference $L_y$. In presence of Neumann B.C. at both ends, the eigenmodes and eigenvalues of the la Laplacian
are
\begin{eqnarray}
 \phi_{n,m}(x,y)&=&\exp\left(2i \pi m\frac{y}{L_y}\right)\cos\left(\pi n\frac{x}{L_x}\right) \\
 \lambda_{n,m}&=&- \left(\frac{2\pi m}{L_y}\right)^2+\left(\frac{\pi n}{L_x}\right)^2 \nonumber \\
 &=& -\left(\frac{2\pi}{L_y}\right)^2\left|m+\tau n\right|^2 \;\;, \tau=\frac{iL_y}{2L_x}
\end{eqnarray}
where $n=0,1,\cdots,\infty$ and $m\in\mathbb{Z}$. This spectrum, with the zero mode omitted, is used to define a generalized zeta-function:
\begin{equation}
 Z(s)=\sum_{\begin{array}{c} n\geq 0,m\in\mathbb{Z} \\
			      (n,m)\ne(0,0)
		\end{array}} \frac{1}{|\lambda_{n,m}|^s}.
\end{equation}
The sum is convergent for $Re(s)>1$ and its analytical continuation to $s=0$ provides a (zeta) regularization for the logarithm of the determinant:
\begin{equation}
 Z'(0)=- \sum_{\begin{array}{c} n\geq 0,m\in\mathbb{Z} \\
			      (n,m)\ne(0,0)
		\end{array}}
		 \ln |\lambda_{n,m}| = -\ln \det{}' \Delta.
\end{equation}
To compute $Z(s)$, we introduce another function
\begin{equation}
 G(s)=\sum_{\begin{array}{c} n,m\in\mathbb{Z} \\
			      (n,m)\ne(0,0)
			      \end{array}}
			      \frac{1}{|n+\tau m|^{2s}},
\end{equation}
such that
\begin{equation}
 Z(s)=\frac{1}{2}\left(\frac{L_y}{2\pi}\right)^{2s} \left( G(s) +2\zeta(2s) \right)
\end{equation}
and
\begin{eqnarray}
  Z'(0)&=& \ln\left(\frac{L_y}{2\pi} \right)\left(G(0)+2\zeta(0)\right) \nonumber\\
    &&+\frac{1}{2}\left(G'(0)+4\zeta'(0)\right) \label{eq:Zp}
\end{eqnarray}
($\zeta(s)=\sum_{n>0} n^{-s}$ is the Riemann zeta-function).
The analytic continuation of  $G(s)$ to $s=0$ is a standard result (see for instance Eq. 4.4 of \cite{osgood_extremals_1988}) :
\begin{eqnarray}
  G(0)&=&-1 \\
  G'(0)&=&-\ln\left((2\pi)^2\left|\eta(\tau)\right|^4\right).
\end{eqnarray}
As for $\zeta$, we have $\zeta(0)=-\frac{1}{2}$ and $\zeta'(0)=-\frac{1}{2}\ln\left(2\pi\right)$.
Plugging these results into Eq.~\ref{eq:Zp}, we  get
\begin{eqnarray}
  Z'(0)&=&-2\ln\left(\frac{L_y}{2\pi} \right) \nonumber\\
  &&-\frac{1}{2}\left(
    \ln\left((2\pi)^2\left|\eta(\tau)\right|^4\right) +2\ln\left(2\pi\right)
  \right)\\
  &=&-\ln\left(L_y^2 \left|\eta(\tau)\right|^2\right).
\end{eqnarray}
We finally obtain :
\begin{eqnarray}
 \ln \det{}' \Delta_{\rm cyl.}&=&\ln\left(L_y^2 \left|\eta(\tau)\right|^2\right)\\
  &=& \ln\left(L_x L_y\right) \nonumber\\
 &&+\ln\left( \frac{L_y}{L_x} \left|\eta\left(\frac{iL_y}{2L_x}\right)\right|^2\right),
\end{eqnarray}
as  announced in Eq.~\ref{eq:det_cyl}.
Note that the $\ln(L_x L_y)$ term in the equation above corresponds to that of Eq.~\ref{eq:det_chi} (with $\chi=0$).

\section{Torus case}
For completeness we  also mention that the method above applies directly to the case of the torus.
In that case the result reads \cite{osgood_extremals_1988}:
\begin{eqnarray}
\ln \det{}' \Delta_{\rm torus}&=& \ln\left(L_x L_y\right) \nonumber\\
 &&+\ln\left( \frac{L_y}{L_x} \left|\eta\left(\frac{iL_y}{2L_x}\right)\right|^4\right).\nonumber
\end{eqnarray}
In terms of $p_{\rm max}$ it  gives (per Nambu-Goldstone mode):
\begin{eqnarray}
 -\ln(p_{\rm max}^{\rm osc,torus}) &=& {\rm const.} N-\frac{1}{4}\ln\left(N\right)\nonumber\\
 &&-\frac{1}{2}\ln\left[\sqrt{\frac{L_y}{L_x}} \left|\eta\left(\frac{iL_y}{2L_x}\right)\right|^2\right].\nonumber
\end{eqnarray}
We finally add the TOS contribution to get:
\begin{eqnarray}
 -\ln(p_{\rm max}^{\rm osc+TOS,torus}) &=& {\rm const.} N+ \frac{1}{4}\ln\left(N\right)\nonumber\\
 && -\frac{1}{2}\ln\left[\sqrt{\frac{L_y}{L_x}} \left|\eta\left(\frac{iL_y}{2L_x}\right)\right|^2\right].\nonumber
\end{eqnarray}

\bibliography{ShannonGoldstone2d.bib}{}
\end{document}